
\newskip\oneline \oneline=1em plus.3em minus.3em
\newskip\halfline \halfline=.5em plus .15em minus.15em
\newbox\sect
\newcount\eq
\newbox\lett
\newdimen\short
\def\adv{\global\advance\eq by1}
\def\set#1#2{\setbox#1=\hbox{#2}}
\def\nextlet#1{\global\advance\eq by-1\setbox
                \lett=\hbox{\rlap#1\phantom{a}}}

\newcount\eqncount
\eqncount=0
\def\equn{\global\advance\eqncount by1\eqno{(\the\eqncount)}}
\def\put#1{\global\edef#1{(\the\eqncount)}           }

\def\mbox#1#2{\vcenter{\hrule \hbox{\vrule height#2in
                \kern#1in \vrule} \hrule}}  
\def\sq{\,\raise.5pt\hbox{$\mbox{.09}{.09}$}\,}
\def\sqb{\,\raise.5pt\hbox{$\overline{\mbox{.09}{.09}}$}\,}

\def\simlt{\mathrel{\lower2.5pt\vbox{\lineskip=0pt\baselineskip=0pt
           \hbox{$<$}\hbox{$\sim$}}}}
\def\simgt{\mathrel{\lower2.5pt\vbox{\lineskip=0pt\baselineskip=0pt
           \hbox{$>$}\hbox{$\sim$}}}}

\def\sig{\sigma}
\def\nabar{{\overline \nabla}}
\def\bR{\overline R}

\def\h1{\hat 1}

\def\cPoly{1}
\def\cDDK{2}
\def\cAM{3}
\def\cAmb{4}
\def\cMig{5}
\def\cAMM{6}
\def\cScm{7}
\def\cDT{8}
\def\cCFL{9}
\def\cAmbp{10}

\magnification=1200
\hsize=6.0 truein
\vsize=8.5 truein
\baselineskip 14pt

\nopagenumbers

\rightline{CPTH-A214.1292}
\rightline{LA-UR-92-4333}
\rightline{December 1992}
\vskip .5truecm
\centerline{\bf {SCALING BEHAVIOR OF QUANTUM FOUR-GEOMETRIES }}
\vskip 1truecm
\centerline{Ignatios Antoniadis}
\centerline{Centre de Physique Th\'eorique}
\centerline{Ecole Polytechnique}
\centerline{91128 Palaiseau, France}
\vskip .4truecm
\centerline{Pawel O. Mazur}
\centerline{Dept. of Physics and Astronomy}
\centerline{University of South Carolina}
\centerline{Columbia, SC 29208}
\vskip .2truecm
\centerline{and}
\vskip .2truecm
\centerline{Emil Mottola}
\centerline{Theoretical Division, T-8}
\centerline{Mail Stop B285}
\centerline{Los Alamos National Laboratory}
\centerline{Los Alamos, NM 87545}
\vskip .5truecm
\centerline{\bf Abstract}
\vskip .4truecm
We propose that large quantum fluctuations of the conformal factor
drastically modify classical general relativity at cosmological distance
scales, resulting in a scale invariant phase of quantum gravity in the far
infrared. We derive scaling relations for the partition function and
physical observables in this conformal phase, and suggest quantitative
tests of these relations in numerical simulations of simplicial four
geometries with $S^4$ topology. In particular, we predict the form of the
critical curve in the coupling constant plane, and determine the scaling of
the Newtonian coupling with volume which permits a sensible continuum
limit. The existing numerical results already provide some evidence of this
new conformal invariant phase of quantum gravity.
\hfill\break
\vfill\eject

\footline={\hss\tenrm\folio\hss}\pageno=1

\par
At accessible distance scales, from centimeters to light years,
gravitational phenomena are described quite accurately by the classical
Einstein
theory. It is very well known that this theory is beset with ultraviolet
divergences at the quantum level and is virtually certain to require
drastic
modification at the Planck scale. A corollary of this severe behavior at
ultra-short distances would seem to be mild behavior in the infrared. One
does
not normally think of quantum fluctuations of the metric field as important
at
large distances, and indeed in perturbation theory around flat space there
is no
sign of infrared problems. However, gravitation is a long range force
characterized by massless excitations that cannot be shielded. If the classical
spacetime is curved on some characteristic distance scale $\ell$,
fluctuations
with wavelengths longer than $\ell$ need not remain small, a fact long ago
pointed out in Newtonian theory by Jeans. Infrared behavior of fluctuations
depends very much on the background geometry, and may force large
modifications
of the classical background.

The quantum analog of the classical background is the ground state. Prior
to and independent of the question of how to tame the ultraviolet
divergences
bedeviling the quantum theory, the question of what is its correct ground
state presents itself. This issue is one that intrinsically involves
infrared
properties, where we may hope to say something sensible in the language of
low energy effective lagrangians, without knowledge of Planck scale
phenomena.

In order to construct a low energy effective lagrangian for gravity that
incorporates infrared fluctuations correctly, one must decide first of all
what is the relevant order parameter field at large distance scales. Here
observational cosmology comes to our aid to suggest that the FRW scale
factor, or more generally, the conformal factor of the metric tensor should
play
an important role. In the classical Einstein theory this scalar part of
the metric does not propagate. It is determined in terms of the matter
sources and has no independent dynamics of its own. At the quantum level
there
is a trace anomaly in the energy-momentum tensor of conformally coupled
matter
fields. The existence of the conformal anomaly means that the classical
constraints which fix the scalar part of the metric fluctuations in terms
of matter sources cannot be maintained upon quantization. In other words,
the
conformal factor becomes {\it unconstrained} in the full quantum theory.

The low energy effective lagrangian must be modified accordingly to take
account of the trace anomaly and fluctuations of the conformal factor, and
this
modified theory reanalyzed to discover the correct infrared behavior of the
quantum theory of gravity. In two dimensional quantum gravity, otherwise
known
as non-critical string theory, the effective action induced by the trace
anomaly modifies the dynamics of the conformal factor at all distance
scales in
dramatic fashion. Certainly nothing like the fractal behavior and scaling
relations of random surfaces in 2D gravity can arise from the classical two
dimensional Einstein-Hilbert action which (being a topological invariant)
yields
no dynamics whatsoever [\cPoly -\cDDK].

In an earlier paper we obtained the effective Wess-Zumino action induced by
the trace anomaly of conformal matter in four dimensions. We analyzed this
continuum effective theory showing that it possesses a non-trivial,
infrared
stable fixed point, characterized by certain anomalous scaling relations
[\cAM].
We argued that this scale invariant phase is the ground state of 4D quantum
gravity approached at distance scales much larger than the horizon length
of any
given classical background. Conformal symmetry, apparently broken by the
trace
anomaly is restored dynamically by the large fluctuations in the conformal
factor
at these scales. An important consequence of restoration of scale
invariance
at large distances is the screening of the effective cosmological term as
measured by the average scalar curvature in the ground state. In
particular, the
two-point correlator of Ricci scalars $\langle R(x) R(x') \rangle$ falls to
zero with a certain universal power of the invariant distance between $x$
and
$x'$ that depends only on the total number of massless fields in the theory
through the effective central charge of the theory.

In principle, it should be possible to verify the existence of such a scale
invariant phase of 4D quantum gravity by appropriate numerical simulations.
In this letter we suggest that a comparison of the continuum predictions
can
be made with simplicial simulations of four-geometries with the topology of
$S^4$ [\cAmb-\cMig]. This topology includes in particular the physically
interesting case of Euclidean de Sitter space. We first derive the scaling
behavior of the partition function for this topology, or equivalently the
behavior of the fixed volume partition function at large volumes $V$. The
continuum theory predicts the existence of a critical curve for the
cosmological
term $\lambda$ as a function of the Newtonian coupling $\kappa$. This
implies
that the corresponding parameters of the dynamical triangulation approach
must
obey a quadratic relation in the scaling limit with the slope of the linear
term
completely determined in terms of pure numbers.

Moreover in order to obtain the continuum limit corresponding to the scale
invariant phase of quantum gravity, the Newtonian coupling of the lattice
theory should be scaled to infinity like $\sqrt V$ for large $V$. An
indication
that this is a proper infinite volume limit to take is that apparently only
in
this limit does the average scalar curvature vanish on the lattice [\cAmb].
A
clear and non-trivial test of the existence of the scale invariant phase is
that
the entropy exponent must be independent of the rescaled Newtonian
parameter.
If this turns out to be the case in the numerical simulations, we will have
at
hand a powerful computational tool for 4D quantum gravity in the far
infrared,
and should be able to {\it measure} the graviton contribution to the
effective
central charge of the conformal theory.

Let us first recapitulate our principal results in the continuum
[\cAM,\cAMM]. We began with the conformal decomposition of the metric
$g_{ab}(x)
= e^{2 \sig(x)} \bar g_{ab}(x)$, with $\bar g_{ab}(x)$ a fixed fiducial
metric.
By consideration of the general form of the trace anomaly for conformal
fields in
four dimensions and taking into account the Wess-Zumino integrability
condition,
we determined the general form of the effective action whose $\sig$
variation is
the trace anomaly. Treating this effective action as the fundamental
quantum
action for the $\sig$ field at large distances and requiring that general
covariance be exactly preserved in the vacuum state of this $\sig$ theory,
we
found that the total trace anomaly of the full theory must vanish. In other
words
the absence of diffeomorphism anomalies in quantum gravity requires that
the
vacuum is a scale invariant conformal fixed point where the beta functions
of all
renormalized couplings are zero.

At the fixed point the effective Euclidean action for $\sig$ reads:
$$
S_{eff} = \int d^4 x \sqrt {\bar g}\Bigl\{ \hbox{$Q^2 \over (4
\pi)^2$}\bigl[
\sig {\overline \Delta_4} \sig + \hbox{$1\over 2$}\bigl(\overline G -
\hbox{$2
\over 3$} \sqb \bR \bigr) \sig \bigr] - \hbox{$3 \over \kappa$} e^{2 \alpha
\sig}
\bigl[\alpha^2 (\nabar \sig)^2 +\hbox{$\bR \over 6$} \bigr] + {\lambda}
e^{4\alpha\sig}\Bigr\}
\equn\put\Seff
$$
where $G$ is the Gauss-Bonnet integrand whose integral is the Euler number,
$$
\chi_{_E} = {1 \over 32 \pi^2} \int d^4 x {\sqrt g}\  G \ ,
\equn\put\euler
$$
and $\Delta_4$ is the Weyl covariant fourth order operator acting upon
scalars:
$$
\Delta_4 = \sq ^2 + 2R^{ab} \nabla_a
\nabla_b - \hbox{$2 \over 3$} R {\sq}+ \hbox{$1 \over 3$}(\nabla^a
R)\nabla_a .
\equn\put\fourth
$$
The quantity $Q^2$ plays the role of the central charge at the infrared
fixed point and is proportional to the coefficient of the Gauss-Bonnet term
in
the quantum trace anomaly.

The physical metric at the conformal fixed point becomes
$$
g_{ab}(x) = e^{2 \alpha\sig(x)} \bar g_{ab}(x),
\equn\put\confdef
$$
with $\alpha$ determined by the condition that the Einstein-Hilbert action
have its canonical scale dimension in this metric. This condition gives a
quadratic equation which fixes $\alpha$ in terms of the central charge:
$$
\alpha = {1 - \sqrt {1 - {4 \over Q^2}} \over {2 \over Q^2}}.
\equn\put\alp
$$
With $\alpha$ determined in this way the condition that the volume
(cosmological) term have dimension 4 requires that the cosmological and
Newtonian couplings satisfy the relation\footnote{$^1$}{Note a
typographical
error in eq. (3.17) of
ref. [\cAM] and the difference in the definition of $\lambda$.},
$$
\lambda = {18 \pi^2 \over \kappa^2}f(Q^2)\ ,\qquad f(Q^2) = {\alpha^2 \over
Q^2}\Biggl[ 1 + {4 \alpha^2 \over Q^2} + {6 \alpha^4 \over Q^4}\Biggr] \ .
\equn\put\rel
$$

All of these results were obtained in the continuum by treating the metric
$\bar g_{ab}$ as fixed. In other words the transverse, traceless sector of
the theory containing the physical spin-$2$ gravitons was neglected
completely.
Our basic hypothesis is that these relations remain true in the infrared
when
the graviton modes are included, up to a possible renormalization of the
value
of $Q^2$. More precisely we assume that integration over the transverse
graviton modes generates an effective action for $\sig$ which, when
expanded in
powers of derivatives, has the same form as {\Seff} but with renormalized
coefficients. This assumption we call ``infrared conformal dominance."

This is not at all unreasonable from a Wilsonian effective action point of
view. Consider the functional integration over transverse gravitons (in
other
words over conformal equivalence classes of metrics), as well as matter
fields,
with both infrared and ultraviolet cut-offs, $\ell$ and $a$. At short
distances,
graviton effects may grow uncontrollably due to the presence of the
dimensionful Newtonian coupling $\kappa$, so that $a$ cannot be taken to
zero in
the effective action. Conversely, at large distance scales, the transverse,
tracefree fluctuations should be expected to become less important, so that
the effective action should be regular as the infrared cutoff $\ell$ is
removed. If this is the case, then an infrared stable renormalization group

fixed point of the effective low energy theory is approached as
$\ell\rightarrow
\infty$. Scale invariance at this fixed point then requires that the low
energy
effective action must be of the form {\Seff} when expanded up to four
derivatives
of $\sig$. The uniqueness of the effective action {\Seff} at the infrared
fixed point is still not sufficient to guarantee that the relations {\alp}
and
{\rel} hold after the integration over the graviton modes. For these
relations
to be unmodified in terms of $Q^2$ one needs to assume that the infrared
fluctuations of the gravitons are subdominant compared to the fluctuations
in the
conformal factor.

In {\Seff}, we have dropped a possible ${\bar C}_{abcd}^2\sig$ term. This
is justified by scale invariance at the fixed point, which requires the
vanishing of the beta function of the Weyl-squared coupling, so that the
coefficient of the ${\bar C}_{abcd}^2$ term in the trace anomaly must
vanish
identically at the fixed point. The same reasoning eliminates the
coefficient of
a local $R^2$ term in {\Seff}, which can be checked explicitly at the level
of
the $\sig$ theory alone.   Then the only unknown parameter in the continuum
theory is the contribution of gravitons to $Q^2$. We have calculated this
contribution to $Q^2$ in perturbation theory in
both the Einstein and Weyl theories and found values around $8$, which lead
to an $\alpha \sim 1.2 \ $[\cAMM].

In order to derive the scaling behavior of the partition function of the
effective $\sig$ theory, subject the $\sig$ field to the constant shift
[\cDDK]
$$
\sig \rightarrow \sig + {\omega \over \alpha}
\equn\put\shift
$$
and use the translational invariance of the integration measure $[{\cal D}
\sig]$ to find:
$$
Z(\kappa , \lambda) \equiv \int [{\cal D} \sig] e^{-S_{eff} [\sig]} = e^{-
{Q^2 \over \alpha} \chi_{_E} \omega} Z(\kappa e^{- 2 \omega}, \lambda
e^{4\omega})\ .
\equn\put\partit
$$
In what follows we restrict to the topology $S^4$ for which $\chi_{_E} =2$.
It is convenient to define also the partition function at fixed volume
$$
Z(\kappa ; V) \equiv \int [{\cal D} \sig] e^{\lambda V - S_{eff} [\sig]} \
\delta\Bigl( \int d^4 x \sqrt {\bar g} e^{4 \alpha \sig} - V \Bigr) \ .
\equn\put\fixvol
$$
Then performing the translation {\shift} above, we
obtain\footnote{$^2$}{Scaling
relations of a similar kind were derived in ref. [\cScm] for conformally
self-dual metrics. However, the physical meaning of the second operator
identified in this work with the ``volume" is unclear to us. The
introduction of
this new operator is also the reason why the critical relation {\rel} was
not
obtained.}
$$\eqalign{
Z(\kappa ; V) &= e^{-2 ({Q^2 \over \alpha} + 2)\omega} Z(\kappa e^{- 2
\omega} ; e^{-4\omega} V)\cr
&= V^{-{Q^2 \over 2\alpha} - 1} {\tilde Z}(\kappa V^{-{1\over 2}})\ ,\cr }
\equn\put\volscal
$$
where in the second line we put $e^{4\omega} \propto V$.

{}From these continuum results we turn now to the numerical simulations. The
numerical method that has proven most fruitful up until now is that of
``dynamical triangulation," a variation of Regge calculus in which
geometries are constructed by gluing together fundamental simplices of
fixed
volume [\cDT]. The four-simplices share common faces with their neighbors
which
are $D-1=3$ dimensional simplices, {\it i.e.} regular tetrahedra of edge
length
$a$. The angle between two tetrahedra faces sharing a triangle is
$$
\theta = \arccos \bigl( {1\over D}\bigr) = 1.3181161
\equn\put\thetaD
$$
in $D=4$ dimensions. The volume of a fundamental $J$-simplex is
$$
V_J = a^J \Omega_J = {a^J\over J!}\sqrt{{J + 1} \over 2^J},
\equn\put\volj
$$
where $a$ is the lattice spacing, so that the total volume of the
simplicial manifold is
$$
\int d^4x\ \sqrt{g} \rightarrow N_4 V_4
\equn\put\totvol
$$
if $N_4$ is the total number of $4$-simplices in the configuration.

As in the Regge approach, space is regarded as flat inside the $D=4$
simplices with all the curvature residing on the $D-2=2$ dimensional
hinges,
{\it i.e.} equilateral triangles with the same fixed edge length. If $n_i$
is
the number of $4$-simplices sharing a given equilateral triangle $i$, then
the
deficit angle $\delta_i$ is given by
$$
\delta_i = 2\pi - n_i \theta \ ,
\equn\put\deficit
$$
and the Einstein-Hilbert action takes the value,
$$
\int d^4x\ \sqrt{g} \ R \rightarrow \sum_i \delta_i V_2 = (2\pi N_2 -10
\theta N_4) V_2,
\equn\put\Ein
$$
where $\sum_i n_i = 10 N_4$ has been used (since each $4$-simplex has $10$
triangles in its boundary).

Dynamics is now specified by giving an action for each simplicial
triangulation $\cal T$ of the form $S({\cal T}) = \sum_J k_J N_J ({\cal
T})$,
with each triangulation otherwise having equal statistical weight. Actually
not
all of the $N_J$ are independent. In order for the simplicial complex to
approximate a continuous manifold the numbers of $J$-subsimplices (for $J =
0,
1, \dots , D$) must satisfy some relations, called Dehn-Sommerville
relations
[\cCFL]. In addition we have the Euler relation,
$$
\sum_{J=0}^4 (-)^J N_J = \chi_{_E} \ .
\equn\put\eulerrel
$$
The net result is that only two of the $N_J$ are independent and the action
may
be taken to be of the form,
$$
S({\cal T}) = -k_2 N_2 ({\cal T}) + k_4 N_4 ({\cal T}) \ .
\equn\put\actDT
$$
Comparing this with the Einstein-Hilbert action, and using the
substitutions
{\totvol} and {\Ein} leads to the following identification of the
simplicial action parameters with those of the continuum,
$$\eqalign{
k_2 &= {{\sqrt 3} \pi \over 4} {a^2 \over \kappa}\cr
k_4 &= {5 {\sqrt 3} \theta \over 4}{a^2 \over \kappa} + {{\sqrt 5}\over 96}
{\lambda a^4} \cr}
\equn\put\param
$$
where the numerical coefficients come from eq. {\volj}.

Since we are interested in the continuum infrared fixed point of the
lattice theory, we do not add any higher derivative couplings to the
action.
These should correspond to irrelevant operators in the infrared, and in any
case
the coefficients of possible $R^2$ and Weyl-squared terms in the action
vanish
at the conformal fixed point. The trace anomaly induced action {\Seff} is
not
to be added to the lattice action {\actDT} either. It is {\it nonlocal} in
the
full metric {\confdef} and should be generated {\it dynamically} by the
quantum
fluctuations of the simplicial geometries, in analogy with the situation in
the two dimensional case.

Because the number of triangulations with fixed $S^4$ topology which can be
made from a given number $N_4$ of $4$-simplices is exponentially bounded
with
respect to $N_4$ [\cAmb-\cMig], the partition function of the dynamical
triangulation approach,
$$
Z_{DT} (k_2 , k_4) \equiv \sum_{{\cal T}} e^{-S({\cal T})} = \sum_{N_4}
Z(k_2 ;
N_4) e^{-k_4 N_4} \simlt \sum_{N_4}  e^{-[k_4 -k_4^c(k_2)] N_4}
\equn\put\partDT
$$
must exist in a region of the coupling constant plane $k_4 > k_4^c(k_2)$.
By
approaching the boundary of this region from above one can hope to arrive
at a
continuum limit in which physical correlation lengths go to infinity when
expressed in lattice units. In other words, one is searching for a critical
curve in the $(k_2 , k_4)$ plane corresponding to a second order phase
transition of the lattice theory defined by {\actDT} and {\partDT}.

The first implication of the continuum $\sig$ theory for the lattice
simulations is that at the infrared fixed point one has the relation {\rel}

which determines the critical curve in the $(k_2 , k_4)$ plane,
$$
k_4^c (k_2) = {5\theta \over \pi} k_2 + {\sqrt 5} f(Q^2) k_2^2 \ ,
\equn\put\kcrit
$$
where use of eqs. {\param} has been made.

This relation should hold for the lattice parameters in the infinite volume
(continuum) limit. In any finite volume simulation there will be an
additive
renormalization of the cosmological term which scales to zero with the
lattice length $a$. Hence the intercept of the curve $k_4^c(k_2)$ will not
vanish in general, for finite volume. However, the slope of the linear term
of
the critical curve {\kcrit} is ${5 \theta \over \pi} = 2.0978469$, a pure
number independent of $Q^2$. Actual simulations with $N_4 \sim 10^4$ seem
to
indicate a critical curve $k_4^c (k_2)$ which is approximately linear with
a
slope slightly more than $2$ [\cAmbp]. The relation {\kcrit} has a
quadratic
term as well which could be used to determine $Q^2$, in principle if the
simulations are run with high statistics and large volumes. However since
$Q^2$
is unknown, and possibly large, and $f(Q^2) \rightarrow 0$ for large $Q^2$,
the
quadratic term could be difficult to measure.

A quite different prediction of infrared conformal dominance and better way
to
measure $Q^2$ is provided by the finite volume scaling relation {\volscal}.
Translating this continuum relation to the lattice by using {\param} we
obtain the following scaling relation for the fixed volume partition
function
at large volumes,
$$
Z(k_2 ; N_4)\sim {\tilde Z}({\tilde k_2}) N_4^{\gamma -3}e^{k_4^c
(k_2)N_4}\ ,
\equn\put\gamcrit
$$
where
$$
{\tilde k_2} = k_2 {\sqrt N_4}\ ,
\equn\put\kscal
$$
and
$$\gamma (Q^2) = 2 - {Q^2 \over 2 \alpha}\ .
\equn\put\gamdef
$$
The content of the scaling relation {\gamcrit} is that, when $k_2$ is
scaled to zero with the square root of the volume as in {\kscal} keeping
${\tilde k_2}$ fixed, the prefactor ${\tilde Z}$ becomes volume independent
and
the entropy exponent $\gamma$ depends only on the effective central charge
$Q$.
Therefore, a clear test of this scaling relation is that $\gamma$ must be
independent of the parameter ${\tilde k_2}$, if our hypothesis of infrared
conformal dominance is correct. Otherwise, $\gamma$ would acquire
non-trivial
$\tilde k_2$ dependence. If $\gamma$ is indeed independent of ${\tilde
k_2}$,
a measurement of $\gamma$ would provide a non-perturbative way to
compute the graviton contribution to the central charge. Note also that the
entropy exponent goes to $-\infty$ in the semiclassical limit $Q^2
\rightarrow
\infty$, while $\gamma = 1$ for $Q^2 = 4$. In the latter case one would
expect
logarithmic behavior in analogy with the $c=1$ case in two dimensions. For
$Q^2 <4$ the exponent $\alpha$ in {\alp} becomes complex and the theory
could
exhibit a phase transition with qualitatively new phenomena. Perturbative
calculations of the graviton contributions [\cAMM] lead to the value
$\gamma\sim -1.3 \ $.

The simplest observable with nice scaling behavior is the average curvature
$\langle R\rangle$ defined by [\cAmb]:
$$
\langle R\rangle \equiv \biggl\langle {\int d^4 x{\sqrt g}R\over\int d^4
x{\sqrt g}}  \biggr\rangle = {2\pi V_2\over V_4}\biggl[{1\over
N_4}{\partial\over\partial k_2}\ln Z(k_2 ;N_4) - {5\theta\over\pi}\biggr] \ ,
\equn\put\avR
$$
where the last equality holds for fixed volume and we used the relations
{\totvol}, {\Ein} and {\actDT}. Now inserting the scaling behavior
{\gamcrit} and the expression {\kcrit} one obtains:
$$\eqalign{
\langle R\rangle &= {2\pi V_2\over V_4}\biggl[2\sqrt{5} f(Q^2) k_2 +
{1\over
{\sqrt N_4}} {\partial\over\partial{\tilde k}_2}\ln {\tilde Z}({\tilde k}_2
)\biggr]\cr  &\sim {1\over\sqrt{N_4}}\rightarrow 0 \ ,\cr}
\equn\put\scalR
$$
where the proportionality factor depends only on $Q^2$ and the rescaled
$\tilde k_2$. This shows that the average curvature scales to zero with the
square root of the volume. Hence scaling $k_2 \rightarrow 0$ with ${\tilde
k}_2$ fixed yields a large volume continuum limit consistent with naive
dimensional analysis. This removes the main obstruction to interpretation
of
the numerical simulations for the continuum theory mentioned in ref.
[\cAmb],
since $\langle R \rangle_{\rm lattice} = a^2 \langle R \rangle_{\rm
continuum}$
vanishes as $k_2 \sim N_4^{-1/2}\rightarrow 0$. This is consistent with the
linear behavior of $\langle R \rangle$ with $k_2$ observed numerically in
ref.
[\cAmb].

Another observable that has been used to search for a continuum limit at a
critical value of $k_2$ is the integrated curvature-curvature correlator,
$$
-{1 \over V}{\partial^2\over \partial k_2^2}\ln Z(k_2 ;N_4)
=  \int d^4 x\ \sqrt{g}\  \langle R(x) R(0) \rangle - V \langle R\rangle^2
\ .
\equn\put\Rcon
$$
If we substitute {\gamcrit} into this expression and use {\kcrit} and
{\kscal}, thereby scaling $k_2\rightarrow 0$ like $N_4^{-{1\over 2}}$, this
quantity goes to a {\it constant} in the large volume limit, in contrast to
the divergent behavior at a finite value of $k_2$ that has been suggested
in
the searches for an ultraviolet fixed point.

Finally we would like to mention that considering only the first term of
the
curvature-curvature correlator in the r.h.s. of {\Rcon} is also interesting
from our present perspective. This is
$$\eqalign{
\int d^4 x\ \sqrt{g}\  \langle R(x) R(0) \rangle
& \sim {1\over N_4} \biggl\langle \Bigl( N_2 - {5 \theta \over \pi}N_4
\Bigr)^2
\biggr\rangle \cr &= {1 \over N_4 Z} \Bigl( {\partial \over \partial k_2} -
{5 \theta \over \pi}N_4 \Bigr)^2 Z(k_2 ; N_4)\ . \cr}
\equn\put\Rcor
$$
If we substitute {\kcrit}, {\gamcrit} and {\kscal} into this last
expression,
a short calculation shows this correlation function also goes to a finite
constant, independent of $N_4$ in the large volume limit. This means that
the correlator $\langle R(x) R(0) \rangle$ must fall off faster than $\vert
x
\vert^{-4}$ for large $\vert x \vert$, in order for the integral over $x$
to converge. Contrast this convergent behavior of {\Rcor}
in the conformal invariant phase of
4D gravity to the behavior of the same quantity in the classical theory. In
that case the trace of the classical Einstein equations, $R=4\Lambda$
rigidly fixes the scalar curvature in terms of the cosmological constant.
Hence {\Rcor} must {\it diverge} linearly with the volume $N_4$ in the
classical
theory, if the cosmological term is non-vanishing.

We conclude that if the quantum fluctuations of the conformal factor
described in this letter really dominate in the far infrared, then any
cosmological term in the continuum theory is screened completely at the
largest
distance scales by these fluctuations, or in other words, the {\it
effective}
cosmological constant in the scale invariant vacuum of 4D quantum gravity
must
vanish identically.

\vskip1.0cm
\centerline{\it Acknowledgements}
\vskip.5cm
We are indebted to J. Ambj{\o}rn for valuable discussions and sharing with
us
his numerical results. This work was supported in part by the NATO grant
CRG900636, and by the EEC contracts SC1-915053 and SC1-CT92-0792. I. A. and
P. O. M, and E. M. thank respectively Los Alamos National Laboratory and
the
Ecole Polytechnique for their hospitality during the completion of this
work.

\baselineskip = 15pt
\parindent=-10 pt
\vfill
\eject

\centerline{\bf REFERENCES }
\vskip 1cm

\item{\cPoly}. A. M. Polyakov, {\it Phys. Lett.} {\bf B103} (1981) 207;
\hfill\break
V. G. Knizhnik, A. M. Polyakov, and A. B. Zamolodchikov, {\it Mod. Phys.
Lett.} {\bf A3} (1988) 819.\hfill\break

\item{\cDDK}. F. David, {\it Mod. Phys. Lett.} {\bf A3} (1988) 1651;
\hfill\break
J. Distler and H. Kawai, {\it Nucl. Phys.} {\bf B321} (1989) 509.
\hfill\break

\item{\cAM}. I. Antoniadis and E. Mottola, {\it Phys. Rev.} {\bf D45}
(1992) 2013. \hfill\break

\item{\cAmb}. J. Ambj{\o}rn and J. Jurkiewicz, {\it Phys. Lett.} {\bf B278}
(1992) 42;\hfill\break
J. Ambj{\o}rn, J. Jurkiewicz, and C. Kristjansen, Niels Bohr Institute
report
NBI-HE-92-53 (Bulletin Board: hep-th@xxx.lanl.gov - 9208032). \hfill\break

\item{\cMig}. M. E. Agishtein and A. A. Migdal, {\it Mod. Phys. Lett.} {\bf
A7} (1992) 1039; Princeton U. report PUPT-1311 (1992). \hfill\break

\item{\cAMM}. I. Antoniadis, P. O. Mazur and E. Mottola, Los Alamos report
LA-UR-92-1483, (Bulletin Board: hep-th@xxx. lanl. gov - 9205015) to appear
in {\it Nucl. Phys.} {\bf B}. \hfill\break

\item{\cScm}. C. Schmidhuber, CalTech report CALT-68-1745
Jun. 1992 (Bulletin Board: hep-th@xxx.lanl.gov - 9112005).\hfill\break

\item{\cDT}. J. Ambj{\o}rn, B. Durhuus, J. Frohlich and P. Orland, {\it
Nucl.
Phys.} {\bf B270} (1986) 457; \hfill\break A. Billoire and F. David, {\it
Nucl.
Phys.} {\bf B275} (1986) 617; \hfill\break D. V. Boulatov, V. A. Kazakov,
I. K.
Kostov and A. A. Migdal, {\it Nucl. Phys.} {\bf B275} (1986) 641.
\hfill\break

\item{\cCFL}. N. H. Christ, R. Friedberg, and T. D. Lee, {\it Nucl. Phys.}
{\bf B202} (1982) 89. \hfill\break

\item{\cAmbp}. J. Ambj{\o}rn, private communication. \hfill\break

\end